# Giant Uncompensated Magnon Spin Currents in X-type Magnets

Zi-An Wang,[1,2] Bo Li,[3,*] Shui-Sen Zhang,[4,2] Wen-Jian Lu,[1] Mingliang Tian,[4,5] Yu-Ping Sun,[4,1]
Evgeny Y. Tsymbal,[6] Kaiyou Wang,[7,8,†] Haifeng Du,[4,9,‡] and Ding-Fu Shao[1,§]

[1] *Key Laboratory of Materials Physics, Institute of Solid State Physics, HFIPS, Chinese Academy of Sciences, Hefei 230031, China*
[2] *University of Science and Technology of China, Hefei 230026, China*
[3] *MOE Key Laboratory for Nonequilibrium Synthesis and Modulation of Condensed Matter, Shaanxi Province Key Laboratory of Quantum Information and Quantum Optoelectronic Devices, School of Physics, Xi'an Jiaotong University, Xi'an 710049, China*
[4] *Anhui Province Key Laboratory of Low-Energy Quantum Materials and Devices, High Magnetic Field Laboratory, HFIPS, Chinese Academy of Sciences, Hefei, Anhui 230031, China*
[5] *School of Physics and Materials Science, Anhui University, Hefei 230601, China*
[6] *Department of Physics and Astronomy & Nebraska Center for Materials and Nanoscience, University of Nebraska, Lincoln, Nebraska 68588-0299, USA*
[7] *State Key Laboratory for Superlattices and Microstructures, Institute of Semiconductors, Chinese Academy of Sciences, Beijing 100083, China*
[8] *Center of Materials Science and Optoelectronics Engineering, University of Chinese Academy of Sciences, Beijing 100049, China*
[9] *Institute of Physical Science and Information Technology, Anhui University, Hefei 230601, China*

[*] libphysics@xjtu.edu.cn; [†] kywang@semi.ac.cn; [‡] duhf@hmfl.ac.cn; [§] dfshao@issp.ac.cn

Magnon spin currents in insulating magnets are useful for low-power spintronics. However, in magnets stacked by antiferromagnetic (AFM) exchange coupling, which have recently aroused significant interest for potential applications in spintronics, these currents are largely counteracted by opposite magnetic sublattices, thus suppressing their net effect. Contrary to this common observation, here, we show that magnets with X-type AFM stacking, where opposite magnetic sublattices form orthogonal intersecting chains, support giant magnon spin currents with minimal compensation. Our model Hamiltonian calculations predict magnetic chain locking of magnon spin currents in these X-type magnets, significantly reducing their compensation ratio. In addition, the one-dimensional nature of the chain-like magnetic sublattices enhances magnon spin conductivities surpassing those of two-dimensional ferromagnets and canonical altermagnets. Notably, uncompensated X-type magnets, such as odd-layer antiferromagnets and ferrimagnets, can exhibit magnon spin currents polarized opposite to those expected by their net magnetization. These unprecedented properties of X-type magnets, combined with their inherent advantages resulting from AFM coupling, offer a promising new path for low-power high-performance spintronics.

Spintronics utilizes a magnetic order parameter of magnetic materials as a state variable for information processing and storage [1]. Spin currents play the key role in spintronics, as they can interact with the magnetic order parameter generating magnetoresistive responses useful for information read-out and exerting spin torques on the magnetic order parameter practical for information write-in [2-8]. Conventionally, spin-current generation is associated with the electric-field-driven motion of electrons which is spin polarized by exchange or spin-orbit interactions. However, this electron motion suffers from the energy dissipation originating from scattering by disorder and/or phonons.

In contrast to electrons carrying both charge and spin, magnons—quantized spin waves, only carry spin angular momentum and thus can form a pure spin current without an accompanying electric current [9-14]. In this regard, magnons have obvious advantages over electrons, especially due to their possible generation in magnetic insulators which substantially reduces the energy dissipation [10, 11]. The magnon spin current can be generated by several external stimuli, such as a temperature gradient [15, 16], a microwave [17], a photoexcitation [18, 19], etc. These approaches have been mostly demonstrated for ferromagnetic (FM) or ferrimagnetic (FIM) materials due to their large magnetization, while the efficiency of the magnon spin current generation is yet to be improved. Recently, giant magnon spin conductivity has been reported for ultrathin FIM films supported by the two-dimensional (2D) confinement [20, 21]. This achievement suggests an efficient route for high-performance magnonic applications using 2D systems such as ultrathin films and/or van der Waals materials. The robustness of the magnetic order in the 2D limit and interface (surface) scattering [22] are the existing challenges of this approach.

Recently, magnets stacked by antiferromagnetic (AFM) exchange coupling have attracted an increasing interest for spintronics due to the spin dynamics being much faster and the stray fields being much smaller than these in FM systems [23-27]. Magnon excitations in AFM-stacked magnets typically occur at much higher frequencies, often in the terahertz range, compared to ferromagnets. However, due to the counteraction of the two magnetic sublattices, the magnon spin currents in AFM-stacked magnets are expected to be strong only for the case of ferrimagnets with large magnetization. For example, a spin Seebeck effect in conventional antiferromagnets can be induced under an applied magnetic field, which only marginally breaks the magnetic compensation [28-32]. A spin Nernst effect may occur even when the chiral modes are degenerate [33-36]. However, this effect primarily originates from the inter-band contributions of certain specific states, which cannot guarantee sizable magnon spin conductivity. Recently, unconventional antiferromagnets, including altermagnets [37-47] and some non-collinear antiferromagnets [48-55], have been proposed to host non-relativistically spin-split band structure supportive to a highly spin-polarized [48,56] or even half-metallic [57,58] Fermi surface capable of generating a strong spin current. However, although these materials support spin-split magnon bands [59-65], this property is not sufficient for strong magnon spin currents. This is because the Bose-Einstein distribution of magnons gathers the contribution from all states of the opposite magnetic sublattices, and hence the magnon spin current might be still largely compensated.

In this letter, we show that the efficient generation of magnon spin currents is possible using the recently discovered antiferromagnets and ferrimagnets with the X-type AFM stacking [66]. Due to the unique cross-chain pattern of the two magnetic sublattices, these X-type AFM-stacked magnets (or X-type magnets for short) support barely compensated magnon spin currents with giant magnon spin conductivities. These results open a new route for the efficient magnon spin-current generation for low-power spintronics.

We start from considering a collinear AFM-stacked magnet with two magnetic sublattices that is described by the Heisenberg Hamiltonian

$$H = -J_{F_A}\sum_i \mathbf{M}_{A_i} \cdot \mathbf{M}_{A_{i+1}} - J_{F_B}\sum_i \mathbf{M}_{B_i} \cdot \mathbf{M}_{B_{i+1}}$$
$$-J_{AF}\sum_{i,j} \mathbf{M}_{A_i} \cdot \mathbf{M}_{B_j} + K_A\sum_i (\hat{\mathbf{n}}\mathbf{M}_{A_i})^2 + K_B\sum_i (\hat{\mathbf{n}}\mathbf{M}_{B_i})^2. \quad (1)$$

Here, the local moment $\mathbf{M}_A$ ($\mathbf{M}_B$) in sublattice A (B) is indexed by $i$ and $j$, $J_{F_A}$ ($J_{F_B}$) > 0 is the nearest-neighbor FM exchange parameter in sublattice A (B), $J_{AF} < 0$ is the nearest-neighbor inter-sublattice AFM exchange parameter, $\hat{\mathbf{n}}$ is the Néel vector, and $K_A$ ($K_B$) is the magnetic anisotropy of sublattice A (B). A temperature gradient $\nabla_m T$ along direction $m$ drives a magnon spin current

$$j_n^s = \sigma_{nm}^s \nabla_m T. \quad (2)$$

In AFM-stacked magnets, the thermal magnon spin conductivity $\sigma_{nm}^s$ can be decomposed into two eigenmodes, $\alpha$ and $\beta$, so that

$$\sigma_{nm}^s = \sigma_{nm}^\alpha + \sigma_{nm}^\beta. \quad (3)$$

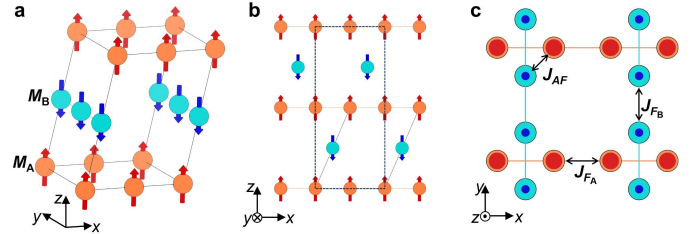

**Fig. 1:** Schematic of a crystal with X-type AFM stacking. (**a**) A primitive unit cell of the crystal. (**b**) Side view of the crystal, where the dashed box denotes the conventional unit cell and the solid box denotes the primitive unit cell. (**c**) Top view of two adjacent layers of the crystal.

For the longitudinal ($n = m$) magnon spin conductivity generated by the spin Seebeck effect, we have

$$\sigma_{nn}^i = \int \frac{d^2\mathbf{k}}{(2\pi)^2}\sigma_{nn}^i(\mathbf{k}) = \frac{\gamma_0^i \tau_0 \hbar}{k_B T^2}\int \frac{d^2\mathbf{k}}{(2\pi)^2}[v_n^i(\mathbf{k})]^2 E^i(\mathbf{k}) f[E^i(\mathbf{k})], (4)$$

where $i \in \{\alpha, \beta\}$, $\gamma_0^\alpha = 1$, $\gamma_0^\beta = -1$, $\tau_0$ is the relaxation time of magnon, $E_k^\alpha$ ($E_k^\beta$) is the magnon band energy at wave vector $k$ contributed by eigenmode $\alpha$ ($\beta$), $v_n^\alpha$ ($v_n^\beta$) is the group velocity along direction $n$, $T$ is temperature, and $f(E) = \frac{e^{E/k_B T}}{(e^{E/k_B T}-1)^2}$.

Since $\sigma_{nn}^\alpha$ and $\sigma_{nn}^\beta$ have opposite sign, a ratio defined by $\xi = \frac{\min\{|\sigma_{nn}^\alpha|,|\sigma_{nn}^\beta|\}}{\max\{|\sigma_{nn}^\alpha|,|\sigma_{nn}^\beta|\}}$ can be regarded as a measure of the magnon spin current compensation by the opposite magnetic sublattices, where $\xi = 0$ implies no compensation, while $\xi = 100\%$ implies full compensation so that $\sigma_{nn}^s = 0$. For example, conventional collinear antiferromagnets with $\mathbf{M}_A = -\mathbf{M}_B$ and degenerate magnon bands exhibit perfect compensation ($\xi = 100\%$) and thus vanishing $\sigma_{nn}^s$. In unconventional collinear antiferromagnets, including altermagnets [37-47] and other non-relativistically spin-split antiferromagnets [58], magnon bands are non-degenerate, allowing for $\xi < 100\%$ and thus for a finite $\sigma_{nn}^s$ [59-61]. However, due to the Bose-Einstein distribution of magnons collecting contributions from all energies (Eq. (4)) including those with zero spin splitting, the resulting $\xi$ may still be close to 100% thus hindering a sizable $\sigma_{nn}^s$. This is also the case for the uncompensated AFM-stacked magnets with $\mathbf{M}_A \neq -\mathbf{M}_B$, i.e. ferrimagnets.

Now, we predict that the X-type AFM stacking can effectively suppress $\xi$, making the thermal magnon spin current barely compensated. The X-type AFM stacking is an extension of the A-, C-, G-, and other types of AFM stacking known for about 70 years [67]. It represents alternating layers of FM-ordered chains parallel to each other within the plane but crossing and AFM-coupled in the adjacent layers. Figure 1 shows a typical structure of the X-type AFM stacking, where A-and B-chains are orthogonal to each other. When $\mathbf{M}_A = -\mathbf{M}_B$, such a system represents a spin-split antiferromagnet with class-I (class-II) X-type AFM stacking for A- and B-chains being symmetric (asymmetric) to each other. Note that there are also spin-degenerate antiferromagnets with class-III X-type AFM stacking where a magnetic unit cell doubles the crystal unit cell [66], which are not discussed in this work.

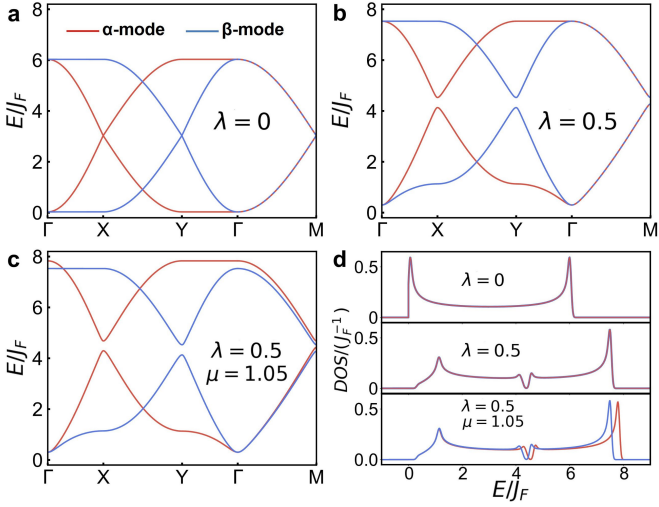

**Fig. 2:** The sublattice-resolved magnon band structure and density of states (DOS) of an X-type AFM bilayer, for magnetic parameters $J_{F_B} = J_F$, $J_{F_A} = \mu J_{F_B}$, $J_{F_B} = J_F$, $J_{AF} = -\lambda J_F$, $K_A = \mu K_B$, and $K_B = 0.01 J_F$. **(a,b)** The magnon band structure of the AFM bilayer with class-I X-type stacking for $\lambda = 0$, $\mu = 1$ (a) and $\lambda = 0.5$, $\mu = 1$ (b). **(c)** The magnon band structure of the AFM bilayer with class-II X-type stacking for $\lambda = 0.5$, $\mu = 1.05$. **(d)** The DOS corresponding to the magnon band structure shown in (a-c).

Figure 1(c) reveals a typical feature of the X-type AFM stacking, namely that the parallel FM chains are well isolated. As a result, the intralayer FM exchange coupling is dominated by the intrachain exchange $J_{F_A}$ and $J_{F_B}$ and the coupling between the parallel chains is negligible. The interlayer AFM exchange coupling $J_{AF}$ connects the adjacent crossing chains and is usually weaker than $J_{F_A}$ and $J_{F_B}$ due to the interchain nearest-neighbor distance being larger than the intrachain distance.

We first discuss X-type antiferromagnets in the 2D limit with only two atomic layers exhibiting the class-I X-type AFM stacking. Such a bilayer crystal represents an altermagnet with the two FM-ordered chains of opposite magnetic sublattices connected by crystal symmetry. Using Hamiltonian (1), we calculate the magnon spectra as described in Supplemental Material [68], assuming $J_{F_A} = J_{F_B} = J_F$, $J_{AF} = -\lambda J_F$, and $K_A = K_B = 0.01 J_F$. Figure 2(a) shows the results in the absence of the AFM coupling between the cross-chains ($\lambda = 0$). In this case, we find that the spin-split magnon band structure is similar to the electronic band structure [66]. Specifically, the bands corresponding to the $\alpha$-mode ($\beta$-mode) are strongly dispersive along the $\Gamma$-X ($\Gamma$-Y) direction, but dispersionless along the $\Gamma$-Y ($\Gamma$-X) direction. This behavior reflects the fact that the $\alpha$-mode ($\beta$-mode) is associated with the A-chain (B-chain) excitations and indicates that the magnon band structure is a superposition of the bands from the isolated A- and B-chains. As a result, the magnon density of states (DOS) shown in Figure 2(d) (top panel) reveals van Hove singularities at the band edges, revealing the one-dimensional (1D) nature of the magnon bands.

As seen from Figure 2(b), including an AFM interlayer exchange coupling in the calculation, $J_{AF} = -0.5 J_F$ ($\lambda = 0.5$), opens a band gap in the magnon spectra. This is accompanied by the bands becoming more (less) dispersive at low (high) energy. At the same time, we observe the increasing low-energy band width due to a finite exchange field applied on both magnetic sublattices (Fig. 2(d), middle panel). Nevertheless, the presence of the low-energy DOS peak even at the sizable $J_{AF}$ implies that the 1D nature of the magnon bands is still sustained.

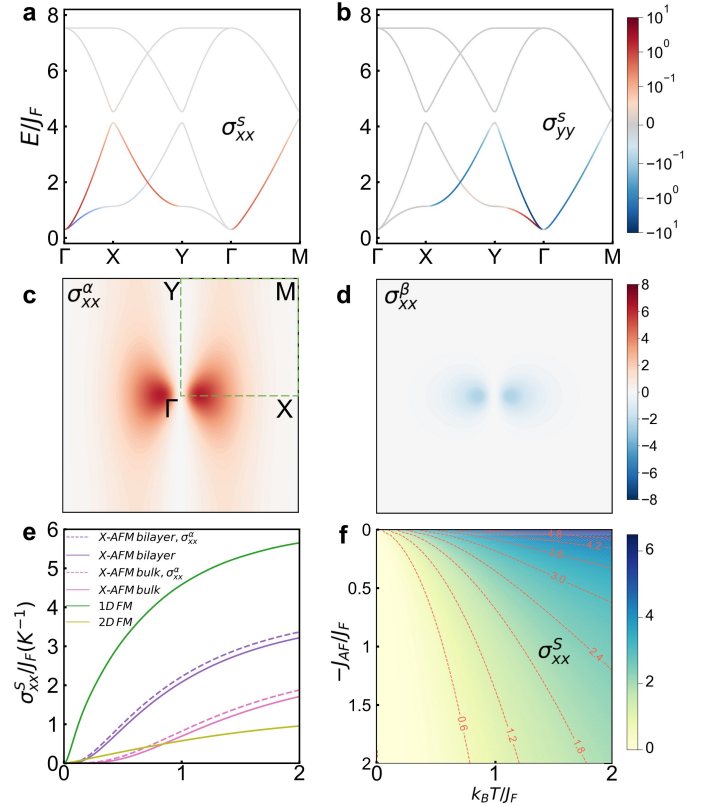

**Fig. 3:** **(a,b)** $\sigma^s_{nn}$ projected magnon bands for an X-type AFM bilayer with $\lambda = 0.5$, $\mu = 1$ for $n = x$ (a) and $n = y$ (b). Other parameters are $k_B T = 0.5 J_F$ and $\tau_0 = 10$ ps. **(c,d)** $k$-dependent $\sigma^\alpha_{xx}(\mathbf{k})$ (c) and $\sigma^\beta_{xx}(\mathbf{k})$ (d) in the 2D Brillouin zone. **(e)** $\sigma^s_{xx}$ as a function of temperature for the X-type AFM bulk and bilayer with $\lambda = 0.5$, $\mu = 1$. The $\sigma^s_{xx}$ for a 1D FM chain and a 2D FM squire lattice are also included for comparison, where the nearest-neighbor FM exchange parameter is $J_F$. We also plot $\sigma^\alpha_{xx}$ of the X-type AFM bilayer as dashed lines for comparison. **(f)** A phase diagram of $\sigma^s_{xx}$ with respect to $J_{AF}$ and $k_B T$.

Such a strong chain-induced band anisotropy can produce a sizable thermal magnon spin current with a minimal $\xi$, provided that the temperature gradient is applied along one of the chain directions. To demonstrate this, we calculate thermal magnon spin conductivities $\sigma_{nn}^S$ driven by temperature gradient $\nabla_n T$ along the A- ($n = x$) or B- ($n = y$) chain. Figure 3(a,b) shows the calculated $\sigma_{xx}^S(\mathbf{k})$ and $\sigma_{yy}^S(\mathbf{k})$ projected on the respective magnon band structures for $\lambda = 0.5$. It is seen that the thermal magnon spin conductivity is dominated by the bands originating from the chains parallel to the temperature gradient. Specifically, for $\nabla_n T$ along the $x$-direction, the $\alpha$-mode (A-chain) controls $\sigma_{xx}^S(\mathbf{k})$ making it largely positive (red color in Fig. 3(a)), while for $\nabla_n T$ along the $y$-direction, the $\beta$-mode (B-chain) controls $\sigma_{yy}^S(\mathbf{k})$ making it largely negative (blue color in Fig. 3(b)). While there is nonvanishing contribution to $\sigma_{nn}^S(\mathbf{k})$ from the chains perpendicular to the temperature gradient (blue color in Fig. 3(a) and red color in Fig. 3(b)), this contribution is small. These features are also seen from the distributions of $\sigma_{xx}^\alpha(\mathbf{k})$ and $\sigma_{xx}^\beta(\mathbf{k})$ in the 2D Brillouin zone (Fig. 3(c,d)), indicating that $\sigma_{xx}^\beta(\mathbf{k})$ is negligible compared to $\sigma_{xx}^\alpha(\mathbf{k})$. Notably, $\sigma_{xx}^\beta(\mathbf{k})$ vanishes along the Γ-Y and Γ-M directions despite the dispersive $\beta$-mode band along these directions (Figs. 2(b) and 3(a,c,d)), clearly reflecting the unidirectional confinement of the chain-like magnetic sublattices. As a result, the sublattice compensation of the thermal magnon spin current is largely suppressed by the X-type AFM stacking. This fact is also evident from Figure 3(e), revealing a very small difference between $\sigma_{xx}^S$ and $\sigma_{xx}^\alpha$ as a function of temperature (violet and magenta dashed and solid lines in Fig. 3(e)). For example, at relatively high temperature $k_B T = 0.8 J_F$, we obtain the magnon spin current compensation ratio of only $\xi = 8\%$. We therefore conclude that the thermal magnon spin current is virtually controlled by a single magnetic sublattice, similar to that for a purely FM system.

The 1D nature of the chain-like sublattices in X-type magnets supports the enhanced magnon spin current due to its unidirectional confinement. We demonstrate this by comparing the calculated $\sigma_{xx}^S$ to the magnon spin conductivities of a 1D FM chain and a 2D FM square lattice, where the nearest-neighbor FM exchange parameters are set to $J_F$ [69]. Although $\sigma_{xx}^S$ of the X-type AFM bilayer is smaller than that of the 1D FM chain, we find that it is much larger than that of the 2D FM square lattice (Fig. 3(e)). This result clearly indicates that the 1D nature of X-type magnets supports giant magnon spin conductivity. Moreover, the $\sigma_{xx}^S$ remains notable even for a much larger $J_{AF}$, as reflected in the diagram of $\sigma_{xx}^S$ with respect to $J_{AF}$ and temperature (Fig. 3(f)). This is also the case for the $\nabla_y T$ along B-chain, where $\sigma_{yy}^S = -\sigma_{xx}^S$ [68].

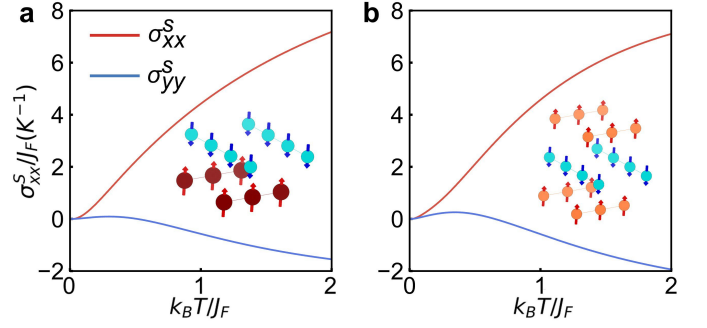

**Fig. 4:** (a) $\sigma_{xx}^S$ and $\sigma_{yy}^S$ as functions of temperature for an X-type AFM trilayer with $\lambda = 0.5$, $\mu = 1$. (b) $\sigma_{xx}^S$ and $\sigma_{yy}^S$ as functions of temperature for an X-type FIM bilayer with $\lambda = 0.5$, $\mu = 1$, and $\mathbf{M}_A = -2\mathbf{M}_B$.

Going from a bilayer to a bulk X-type antiferromagnet with class-I AFM stacking, we find that $\sigma_{xx}^S$ is somewhat reduced (Fig. 3(e)) [68]. This is because each FM-ordered layer in the bulk has two adjacent layers (top and bottom) AFM-coupled to the former, while only one is present in the bilayer. As a result, the effective AFM exchange field on each magnetic sublattice in the bulk is twice that in the bilayer. This enhances the contribution to $\sigma_{xx}^S$ from the chains perpendicular to the temperature gradient. Nevertheless, we still observe weak compensation (e.g. $\xi = 20\%$ at $k_B T = 0.8 J_F$) and sizable $\sigma_{xx}^S$ which is larger than that of the 2D FM square lattice at not too low temperature (Fig. 3(c)).

Similar magnon properties occur in class-II X-type antiferromagnets. In these crystals, A- and B-chains are asymmetric due to slightly different intra-(inter-) chain distances or relative rotations (distortions) of the structural motifs [66]. This is simulated by fixing $J_{F_B}$ and $K_B$ and setting $J_{F_A} = \mu J_{F_B}$ and $K_A = \mu K_B$ with $\mu = 1.05$. We find a similar band structure, except the degeneracy being removed at the Γ point [70], which is a typical feature of a non-altermagnetic spin splitting [58]. This leads to a non-degenerate sublattice-resolved magnon DOS with the peaks at the band edges remaining intact (Fig. 2(c,d)). This results in slightly different $\sigma_{xx}^S$ and $\sigma_{yy}^S$, which stay sizable and uncompensated, as shown in Fig. S2 [68].

We further confirm the advantage of the X-type AFM stacking for magnon spin current generation by considering two uncompensated magnets with sizable net magnetizations. The first one is a class-I X-type AFM trilayer. The second one is a bilayer ferrimagnet where the orthogonal A- and B-chains host non-equivalent moments $\mathbf{M}_A = -2\mathbf{M}_B = 2\mathbf{M}_0$, representing a class-II X-type AFM stacking. Such systems can be prepared by layer-by-layer deposition using modern film-growth techniques [71, 72]. In both cases, the effective exchange fields on the sublattices are different from those in X-type antiferromagnets discussed above, which significantly influence the band structure (Fig. S4). Despite the difference, the basic feature of the X-type

AFM stacking is sustained, i.e. the bands dominant by A-chain (B-chain) excitations are strongly dispersive along the Γ-X (Γ-Y) direction but weakly dispersive along the Γ-Y (Γ-X) direction (Fig. S4). As a result, $\sigma_{xx}^s$ ($\sigma_{yy}^s$) remain majorly contributed by the A-chains (B-chains) [73]. As seen from Figures 4, $\sigma_{xx}^s$ is positive and much larger in these systems compared to the X-type AFM bilayer (Fig. 3(c)), due to the net magnetic moments on the sublattice A being much larger than on the sublattice B. Notably, we find a negative $\sigma_{yy}^s$ for both systems, indicating that a magnon spin current is polarized opposite to that expected by the net magnetic moment $\boldsymbol{M}_0$. This is in contrast to the common expectation and serves as another evidence of the uncompensated magnon spin current nearly locked to the selected directions by the X-type AFM stacking.

When the temperature gradient $\nabla_m T$ is along direction $m$ making angle $\theta$ relative to the $x$-axis, a transverse magnon spin current in X-type magnets appears along the direction $n \perp m$ with conductivity $\sigma_{nm}^s = (\sigma_{xx}^s - \sigma_{yy}^s) \sin\theta \cos\theta$, representing a spin Nernst effect. The maximum $|\sigma_{nm}^s| = |(\sigma_{xx}^s - \sigma_{yy}^s)/2|$ occurs for $m$ along the diagonal directions (Fig. S5) [68]. Since $\sigma_{xx}^s$ and $\sigma_{yy}^s$ have opposite signs and are both large, we expect that the transverse magnon spin current driven by the X-type AFM stacking will be much stronger than that originating from the Berry curvature.

We emphasize that the key property of the X-type AFM stacking resulting in the predicted giant uncompensated magnon spin current is the strong intra-sublattice FM exchange coupling in one direction and negligible in the perpendicular direction. This property is the direct consequence of the chain-like sublattice structure of the X-type AFM stacking producing strong anisotropy of the two magnetic sublattices. The sublattice anisotropy has implications for understanding the thermal magnon spin current generation in AFM-stacked magnets. For example, rutile altermagnets do not exhibit sufficiently strong in-plane structural anisotropy within the magnetic sublattice, resulting in the strong compensation of the magnon spin current (e.g. $\xi = 87\%$ for $RuO_2$ at 300 K) [60,61,68]. In other altermagnetic candidates, such as $M_2X_2O$ ($M$ = V, Cr; $X$ = Se, Te) [43,59], the intra-sublattice nearest neighbors are connected by the M-X-M bonds in one direction and the M-O-M bonds in the perpendicular directions. This sublattice anisotropy makes the FM exchange constants along these directions very different, supportive to a not-too-large $\xi$ and thus a sizable magnon spin current [59,68]. The X-type magnets can be regarded as the extreme case of the AFM-stacked magnets with the strongest structural anisotropy of magnetic sublattices. For example, in the case of typical X-type antiferromagnet $\beta$-$Fe_2PO_5$ [66], which is described by the magnetic parameters $J_{F_A} = J_{F_B} = 32$ meV, $J_{AF} = -9.2$ meV, and $K_A = K_B = -0.44$ meV, we calculate tiny $\xi = 6\%$ and giant $\sigma_{xx}^s = -\sigma_{yy}^s = 30.8$ meV/K at 300 K [68]. This implies an outstanding efficiency of the magnon spin current generation of this X-type antiferromagnet compared to the known AFM-stacked magnets.

Overall, X-type magnets provide conditions for the efficient magnon spin current generation which is usually difficult in antiferromagnets and ferrimagnets. Due to the unique orthogonal cross-chain structure, the counteraction of the two magnetic sublattices in these magnets is effectively suppressed, supporting magnon spin currents with minimal compensation. The 1D nature of the isolated FM chains significantly enhances the magnon spin conductivity and, in combination with the minimal sublattice compensation, makes it giant. These unprecedented properties of the X-type magnets are promising for applications in low-power high-performance spintronics.

**Acknowledgments.** This work was supported by the National Key R&D Program of China (Grant No. 2021YFA1600200), the National Natural Science Funds for Distinguished Young Scholar (Grant No. 52325105), the National Natural Science Foundation of China (Grants Nos. 12274411, 12241405, and 52250418), the Basic Research Program of the Chinese Academy of Sciences Based on Major Scientific Infrastructures (Grant No. JZHKYPT-2021-08), and the CAS Project for Young Scientists in Basic Research (Grant No. YSBR-084). E.Y.T acknowledges support from UNL's Grand Challenges catalyst award. Computations were performed at Hefei Advanced Computing Center.